# Insulator-metal transition shift related to magnetic polarons in $La_{0.67-x}Y_xCa_{0.33}MnO_3$

G. Li, S. -J. Feng, F. Liu, Y. Yang, R. -K. Zheng, T. Qian, X. -Y. Guo, X. -G. Li*

Structure Research Laboratory, Department of Materials Science and Engineering, University of Science and Technology of China, Hefei 230026, P. R. China

## Abstract

The magnetic transport properties have been measured for $La_{0.67-x}Y_xCa_{0.33}MnO_3$ ($0 \le x \le 0.14$) system. It was found that the transition temperature $T_p$ almost linearly moves to higher temperature as $H$ increases. Electron spin resonance confirms that above $T_p$, there exist ferromagnetic clusters. From the magnetic polaron point of view, the shift of $T_p$ *vs*. $H$ was understood, and it was estimated that the size of the magnetic polaron is of 9.7~15.4 Å which is consistent with the magnetic correlation length revealed by the small-angle neutron-scattering technique. The transport properties at temperatures higher than $T_p$ conform to the variable-range hopping mechanism.

* Corresponding author, E-mail: lixg@ustc.edu.cn

## 1. Introduction

Manganese perovskites based on the compound $LaMnO_3$ are attracting considerable theoretical and technological interest by virtue of their unusual magnetic and electronic properties.[1-4] Most notable of these properties is the extremely large change in resistivity that accompanies the application of a magnetic field near ferromagnetic ordering temperature $T_C$, an effect known as 'colossal' magneto-resistance (CMR).

Although the electron-lattice interaction arising from the dynamic Jahn-Teller distortion is considered to be important for the understanding of overall trends of CMR phenomena[5] besides the conventional double-exchange (DE) mechanism,[6] the lattice polaron formation is incomplete to explain the transport properties in connection with the observed CMR effect.[7-9] It is suggested that the $e_g$ carriers of $Mn^{3+}$ ions can be trapped by spin-disorder scattering due to the local deviations of the ferromagnetic surroundings, resulting in the formation of magnetic polarons in the vicinity of $T_C$.[10, 11] A lot of theoretical works[12-16] have demonstrated that the magnetic polaron formation plays a crucial role in CMR phenomena in paramagnetic state. At the same time, many experiments have been devised to understand their unique magnetic features in the paramagnetic state. For instance, the high-temperature inverse susceptibility for $La_{0.7}Ca_{0.3}MnO_3$ is smaller than that expected from Curie-Weiss law at low magnetic field $H < 0.1$ T.[17, 18] Moreover, using a combination of volume thermal expansion (with and without an applied field), magnetic susceptibility and small-angle neutron scattering measurements, De Teresa *et al.*[19] presented direct evidence for the existence of magnetic polarons above $T_C$. They detected the spontaneous formation of localized similar to 12 Å magnetic polarons above $T_C$ which, on application of a magnetic field, grow in size but decrease in number in the compounds $(La_{1-x}A_x)_{2/3}Ca_{1/3}MnO_3$ (where A is Y or Tb). There has also been an increasing realization that manifests the importance of the magnetic polarons in the paramagnetic state of the doped perovskite manganites.

The ferromagnetic ordering temperature $T_C$ of $La_{0.67-x}Y_xCa_{0.33}MnO_3$ system is close to the corresponding insulator-metal transition temperature $T_p$, and both of them decrease linearly with the Y content.[20] In this paper, we selected $La_{0.67-x}Y_xCa_{0.33}MnO_3$ system to study the magnetic field-dependent insulator-metal transition and transport properties above $T_p$. It was found that $T_p$ almost linearly moves to higher temperature with $H$ at a rate of 6~10 K/T. From the magnetic polaron point of view, the shift of $T_p$ *vs.* $H$ was understood. From the rate of the shifts, the magnetic polaron in the vicinity of $T_p$ was estimated to be about 9.7~15.4 Å in size that is consistent to the magnetic correlation length revealed by the small angle neuron scattering (SANS). [19]

## 2. Experiments

The four samples with the nominal compositions $La_{0.67-x}Y_xCa_{0.33}MnO_3$ ($x$ = 0, 0.07, 0.10, and 0.14) were prepared by a solid-state reaction method from starting materials $La_2O_3$, $Y_2O_3$, $MnCO_3$, and $CaCO_3$ of high-purity (>99%). The starting materials were completely mixed, ground, and calcined at 1000°C for 15 hours and 1100°C for 18 hours leading to powders. Then the powders were pressed into pellets and sintered at 1200°C for 24 hours in air with intermediate grindings. Finally, the pellets were again ground to powders, pressed into bars, and sintered at 1250°C for 15 hours. X-ray diffraction (XRD) patterns were recorded by MacScience MAXP18AHF diffractometer using Cu Kα radiation. The temperature dependencies of resistivity for the samples were measured in magnetic fields up to 8 T supplied by Oxford superconducting magnets upon warming from 4.2 K to room temperature by a four-probe technique. ESR experiments on the $La_{0.6}Y_{0.07}Ca_{0.33}MnO_3$ in the temperature range from 160 to 300 K were carried out on a Bruker ER-200D-SRC spectrometer in X-band at the frequency of 9.47 GHz.

## 3. Results and discussion

Figure 1 presents the XRD spectra of the samples at room temperature. No structural change due to the Y doping was detected from the diffraction. Detailed analysis shows that the samples are single phased and well crystallized with a nearly cubic structure, as shown in Fig.1(b). With Y doping, the diffraction peaks move to high angles, denoting that the lattice parameter and volume decrease systematically with *x*, see the insets of Fig.1(a) and (b). This is consistent to the fact that the radius of $Y^{3+}$ ion is smaller than that of $La^{3+}$ ion.

The field-dependent resistivity $\rho$ as a function of $T$ is presented in figure 2 for the four samples. One can see that in the absence of $H$, the insulator-metal transition temperatures $T_p$ ($H$ = 0) move to low temperatures as Y doping increases from 0 to 0.1, and when $x$ reaches 0.14, the transition disappears. At the same time, the magnitude of the zero-field resistivity increases with $x$. The reason [20] is that the average radius of A site ($ABO_3$ perovskite structure) $\langle r_A \rangle$ reduces and $Mn^{3+}$-$O^{2-}$-$Mn^{4+}$ angle $\theta$ deviates from 180°, which will hamper the $e_g$ electrons transferring between the neighboring $Mn^{3+}$ and $Mn^{4+}$ ions. Hence $T_p$ decreases and the resistivity increases with the Y doping. As $H$ increases up to 2

T or larger, an insulator-metal transition appears for $x$ = 0.14 sample. Similar results were also observed in the $Pr_{0.7}Ca_{0.3}MnO_3$ sample.[21] The $<r_A>$ of the both samples is close to 1.18 Å that locates the critical cross-point of paramagnetic insulating, ferromagnetic insulating, and ferromagnetic metallic phases in the phase diagram of temperature *vs.* $<r_A>$ and tolerance factor.[22] The insulating state at the critical point is not stable and can easily be induced to a metallic state by a strong enough $H$.

The magnetic field $H$ can not only induce the insulator-metal transition but also push the transition towards high temperatures and decrease the resistivity. From figure 2, one can easily find that as $H$ increases, $T_p$ for each sample moves to high temperatures and the resistivity in the vicinity of $T_p$ is greatly reduced, leading to CMR effect.

Figure 3 depicts the relationship between $T_p(H)$ and $H$. One may notice that $T_p(H)$ almost linearly moves to high temperatures as $H$ increases, and the increment of $T_p(H)$ per tesla is about 6 ~10 K (8, 7.4, 9.3, and 6.4 K/T for $x$ = 0, 0.07, 0.1 and 0.14, respectively ). Such an increasing rate was also observed in the other CMR manganese oxides.[21, 23] Theoretical calculations predicted that a magnetic field could push the insulator-metal transition to high temperature.[24] We thought there must be a common physical origin underlying for this shift.

From the energy point of view, the external magnetic filed can bring extra energy to the system. The energy can be either proportional to $H^2$, the self-energy of magnetic field; or the interaction energy $E_H$ of the magnetic moments of an entity in the magnetic field. For a magnetic system, it is physically reasonable to regard that the latter form of energy that the magnetic field brings to the system will appear. The latter can be written as

$$E_H = \gamma\mu_B S_{eff} H \ , \qquad (1)$$

in which $\gamma$, $\mu_B$, and $S_{eff}$ are the gyromagnetic ratio, Bohr magneton, and the effective spin of the magnetic moments of the entity, respectively. $\gamma$ is constant and nearly equal to 2 for the manganites due to the frozen orbital.

More and more theoretical calculations[25] and experimental results[19, 26-29] indicated that the local entities—short-range ferromagnetic clusters form in the vicinity of $T_p$. The short-range ferromagnetic clusters can trap charge carriers forming the magnetic polarons. With the application of a magnetic field, the potential energy of the magnetic polarons decreases in the vicinity of $T_p$. As a result, some magnetic polarons will grow in size and merge into the larger long-range ferromagnetic clusters, in which the magnetic polarons liberate/delocalize the self-trapped charge carriers. So, it brings about a drop in resistivity (several orders of magnitude) and the insulator-metal transition ($T_p(H)$) moves to high temperatures. When $H$ becomes larger, the magnetic polarons at higher temperatures will become large in size and also melt into the ferromagnetic metallic phase. In fact, there is a competition between the $H$-induced magnetic polaron merging and thermal fluctuation that will counteract the merging. So, it is reasonably regarded that the increment of $T_p(H)$ per tesla is

$$\delta T_p = \delta E_H / k_B = \gamma\mu_B S_{eff} \delta H / k_B \, . \qquad (2)$$

Here $k_B$ is the Boltzmann constant. From Eq. 2 one can get

$$S_{eff} = \frac{k_B}{\gamma\mu_B} \frac{\delta T_p}{\delta H} \, . \qquad (3)$$

As mentioned above, here $\frac{\delta T_p}{\delta H}$ = 6 ~ 10 K/T. After simple calculation, one can get that $S_{eff}$ = 4.5 ~ 7.4. For $La_{0.67-x}Y_xCa_{0.33}MnO_3$ system, the ratio of $Mn^{3+}/Mn^{4+}$ is 2:1, and the spin value $S$ of the $Mn^{3+}$—$O^{2-}$—$Mn^{4+}$ individual is 1.83 per unit cell.[25] So, the ferromagnetic cluster should consist of a few $Mn^{3+}$—$O^{2-}$—$Mn^{4+}$ individuals. The number $n$ of the individuals that the entity contains equals $S_{eff}/S$ = (4.5 ~ 7.4)/ 1.83 = (2.5 ~ 4). The size of the ferromagnetic surroundings is of the order of 2.5 ~ 4 lattice spacing distances. With the lattice parameter $a$ = 3.86 Å the size of the cluster is estimated to be $na$ = 9.7 ~15.4 Å. The value is in good agreement with the size of magnetic polaron revealed by the recent SANS measurements[19] and the theoretical simulations. [25] One may find that the most important effect induced by the external magnetic field is that $T_p$ moves to a higher temperature due to the magnetic polarons condensed into the metallic phase. $T_p$ moving to a higher temperature means that the external magnetic field increases the long-range ferromagnetic metallic fraction. Thus the CMR effect manifestly appears around $T_p$.

In order to confirm that there are short-range ferromagnetic clusters existing in the paramagnetic matrix, we employed electron spin resonance (ESR) technique to check the magnetic behavior as a function of temperature. For instance, the $La_{0.6}Y_{0.07}Ca_{0.33}MnO_3$ sample, whose $T_p$ is 163 K in the absence of magnetic field, has been selected to perform the ESR experiment in the temperature range from 160 K to 300 K. Figure 4 presents the ESR spectra of the sample at different temperatures. It can been seen at temperatures higher than 240 K, each spectrum consists of a single line with the magnetic resonance field at 3400 Gauss denoted by $S_1$. This resonance can be attributed to the paramagnetic signal.[30] When temperature decreases to 240 K or below, however, there is an extra resonance signal denoted by $S_2$ at lower magnetic field region appears besides $S_1$. Note that the short-range ferromagnetic clusters appear at temperatures much higher than $T_p$. The $S_2$ signal suggests the appearance of short-range ferromagnetic clusters in the matrix of paramagnetic state.[30] The position of $S_2$ moving to low magnetic field means that he ferromagnetic clusters grow in size as the temperature decreases. Similar behaviors were also observed in the remaining $La_{0.67-x}Y_xCa_{0.33}MnO_3$ ($x$ = 0, 0.1) samples which have the insulator-metal transition in the absence of magnetic field, not shown here.

Although the electron-lattice interaction arising from the dynamic lattice Jahn-Teller distortion is considered to be very important for understanding of overall trends of CMR phenomena,[5] the lattice polaron is incomplete to explain the transport properties in connection with the observed CMR. [7-9] Above $T_P$(or $T_C$), the samples are paramagnetic semiconductors, and the conductivity is supposed to emerge from hopping of magnetic polarons due to the random magnetic potential originating from the distribution of inhomogeneous magnetic clusters (magnetic polarons)[27,31] It is suggested that the $e_g$ carriers of $Mn^{3+}$ ions can be trapped by spin-disorder scattering due to the local deviations of the ferromagnetic surroundings, resulting in the formation of magnetic polarons in the vicinity of $T_C$.[10, 11] The self-trapped states of magnetic polarons are theoretically proved to only disappear at $T\rightarrow\infty$.(Nagaev, Ref. 32) The hopping magnetic polarons are widely used to theoretically or experimentally interpret the conductive properties in the paramagnetic state of manganites.[27, 32] Given that $La_{0.67-x}Y_xCa_{0.33}MnO_3$ contains magnetic polarons embedded in the paramagnetic insulating matrix when $T > T_p$, one may propose a variable-range-hopping conduction: $\rho = \rho_0 \exp(T_0 / T)^{1/4}$, where $\rho_0$ is the prefactor and $T_0$ is the characteristic temperature. $T_0 \propto 1/\xi^3 k_B N(E_F)$, in which $\xi$ is the localization length of the trapped charge carriers, $N(E_F)$ is the density of localized states at Fermi level.[33] The plots of ln$\rho$ *vs.* $T^{-1/4}$ are presented in the figure 5. A good linear ln$\rho \sim T^{-1/4}$ relationship at temperatures higher than $T_p$ can be seen for all the Y-doped samples in the presence or absence of magnetic fields. One may find that $T_0^{1/4}$ increases with Y doping meaning that the localization of charge carriers is strengthened. Note that for the Y-undoped sample, there are not enough data recorded for the ln$\rho$ *vs.* $T^{-1/4}$ plots because the $T_p$ is close to the highest temperature (300 K) in our measurements and the plots of ln$\rho \sim T^{-1/4}$ are not presented. Similar transport properties above $T_p$ were also observed in $A_{0.7}B_{0.3}MnO_3$ compounds with different *A* and *B* cations like $La^{3+}$, $Pr^{3+}$, $Nd^{3+}$ and $Ca^{2+}$, $Sr^{2+}$ and $Ba^{2+}$, respectively.[34]

For the transport above $T_p$, where ln$\rho \propto (T_0/T)^{1/4}$ scaling holds, we first made use of the approximations[32] without considering a random potential of mainly magnetic inhomogeneity (ferromagnetic clusters) above $T_P$,

$$k_B T_0 = 18/\xi^3 N(E_F) \tag{4}$$

and

$$R = \{9\xi/8\pi N(E_F) k_B T\}^{1/4} \tag{5}$$

to evaluate the values of $\xi$ and the average hopping distance *R*. The values of $\xi$ range from 0.3 to 0.8 Å, and *R* from 3.3 to 4.1 Å at room temperature for the current system, respectively, by considering the electronic density of states $N(E_F) = 4\times10^{28}$ m$^{-3}$ eV$^{-1}$ revealed by heat capacity measurements[13] and $T_0^{1/4}$ ranging from 56 and 111.8 K$^{1/4}$, see figure 5. Since the localization must exceed the Mn-Mn distance and the hopping distance should be several times greater, these numbers are incompatible with the conventional variable-range hopping. In order to answer the incompatibility, it is proposed that a random potential of mainly magnetic origin is responsible for carrier localization above $T_p$.[32] Thus, the density of states is rectified to be about $9\times10^{26}$ $m^{-3}$ $eV^{-1}$ and $k_B T_0 = 171 U_m v/\xi^3$, where $U_m$ (~ 2 eV) and $v$ (= $a^3$ = $5.7\times10^{-29}$ $m^3$) are the splitting energy between spin-up and spin-down $e_g$ band and the lattice volume per manganese ion, respectively. The corresponding values of $\xi$ now range from 0.8 to 2.1 Å, and $R$ from 10.6 to 13.4 Å at room temperature for the current system, respectively. The numbers are physically plausible since the localization length of $e_g$ electron exceeds the ionic radius of $Mn^{3+}$ and the hopping distances are 3-4 times the Mn-Mn separation. It should be noted that the hopping distances are consistent with the size of magnetic polarons revealed from the $T_p$ shift. It is reasonable to consider that the dimension of a hopping taking place is only the size of a magnetic polaron and the localization of $e_g$ electron trapped by the magnetic polaron is very strong. The field dependence of $\xi$ and $R$ for $La_{0.67-x}Y_xCa_{0.33}MnO_3$ ($0.07 \le x \le 0.14$) system are plotted in figure 6.

The parameter $T_0^{1/4}$ decreases as $H$ increases denoting that the localization length $\xi$ and $R$ increases as $H$ increases and the ability of the magnetic polarons to trap $e_g$ carriers weakens, because with the increase of $H$ more and more magnetic polarons at high temperature would melt into the ferromagnetic metallic phase.

## 4. Conclusions

The transport properties of $La_{0.67-x}Y_xCa_{0.33}MnO_3$ ($0 \le x \le 0.14$) system were measured. The samples with Y content $x \le 0.10$ show an insulator-metal transition in the absence or presence of magnetic field while the $x$ = 0.14 sample shows the transition only if the magnetic field is up to 2 T or larger. The transition temperatures almost move linearly to high temperatures at a rate 6 ~ 10 K/T. Based on the scenario of magnetic polarons existing in the paramagnetic insulating matrix, the shift of the transition and CMR effect were explained. Furthermore, the size of the magnetic polarons was estimated from the shift rate to be 9.7~15.4 Å which is in good agreement to that revealed by the SANS technique and theoretical calculations.

**Acknowledgements**

This work was supported by the Chinese National Natural Science Foundation and the Ministry of Science and Technology of China.

**Figure Captions**

Fig.1 XRD patterns for the $La_{0.67-x}Y_xCa_{0.33}MnO_3$ ($0 \leq x \leq 0.14$) system (a) and the lattice parameter and lattice volume (b). Insets of (a) and (b) show both the peak shift to high angle and lattice volume with *x*, respectively. The dotted lines are guide to eyes.

Fig.2 Transport properties of the $La_{0.67-x}Y_xCa_{0.33}MnO_3$ ($0 \leq x \leq 0.14$) system in different magnetic fields.

Fig.3 Field dependence of the insulator-metal transition temperature $T_p$. It can be seen that $T_p$ almost linearly moves to higher temperature at a rate of 6 ~ 10 K/T. The solid lines are guides to eyes.

Fig.4 ESR spectra in the temperature range from 160 K to 300 K for the $x = 0.07$ sample. Below 240 K, there is an extra signal denoted by $S_2$ appearing besides the paramagnetic signal denoted by $S_1$, which suggests that there are ferromagnetic clusters appearing in the paramagnetic background when $T < 240$ K. Note that the resonance position of $S_2$ shifts gradually to the low field region as *T* decreases.

Fig.5 (a) Plots of $\ln\rho$ *vs.* $T^{-1/4}$ and (b) relationship between $T_0^{1/4}$ and *H*. The dotted lines are guide to eyes.

Fig.6 Field dependence of the localization length $\xi$ (a) and the hopping distance *R* (b) evaluated at room temperature. The open symbols represent the rectified values of $\xi$ and *R*. from Eq.4; the solid symbols depict the values of $\xi$ and *R* without modification from Eq.5.

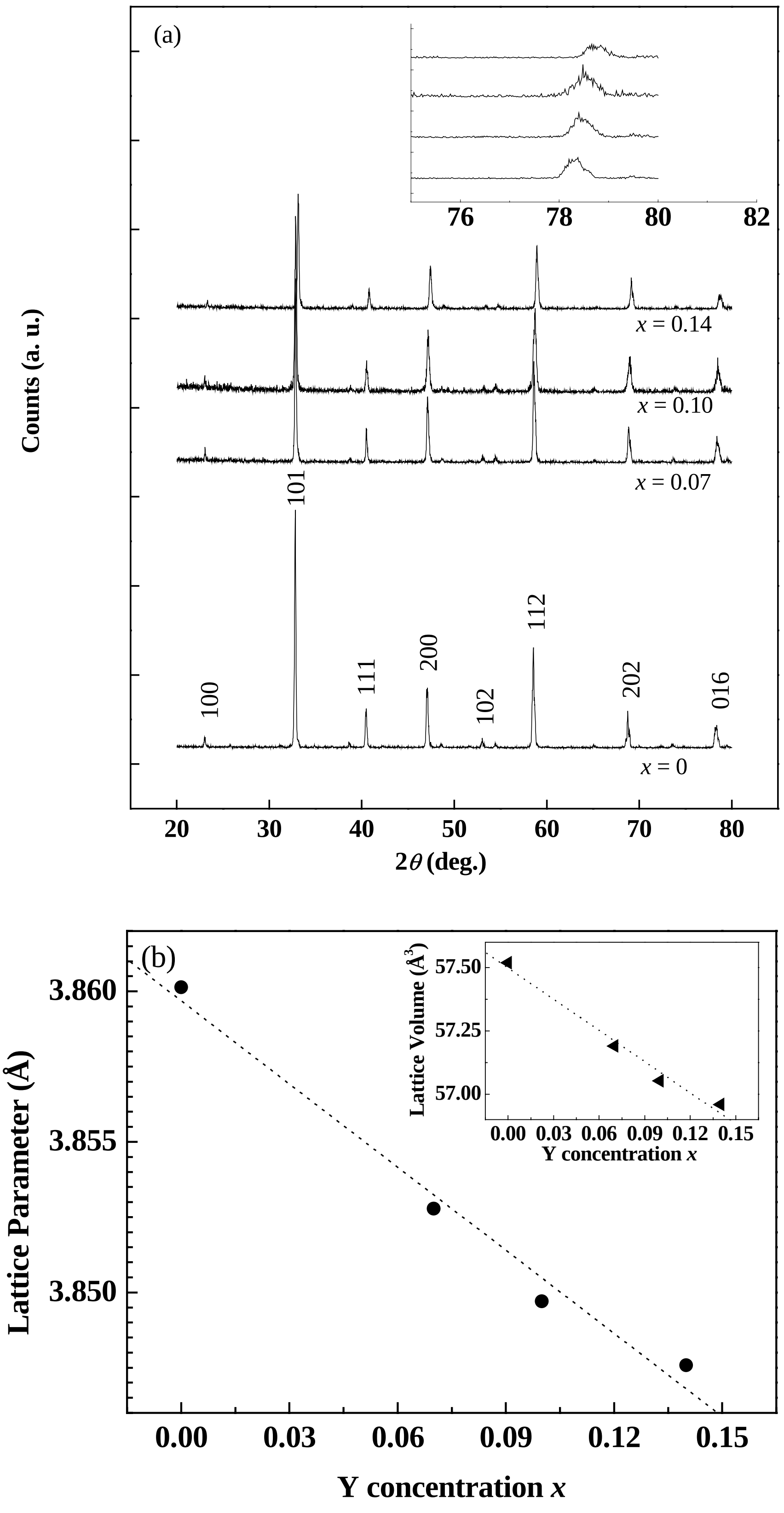

Fig. 1 by G. Li *et al.*

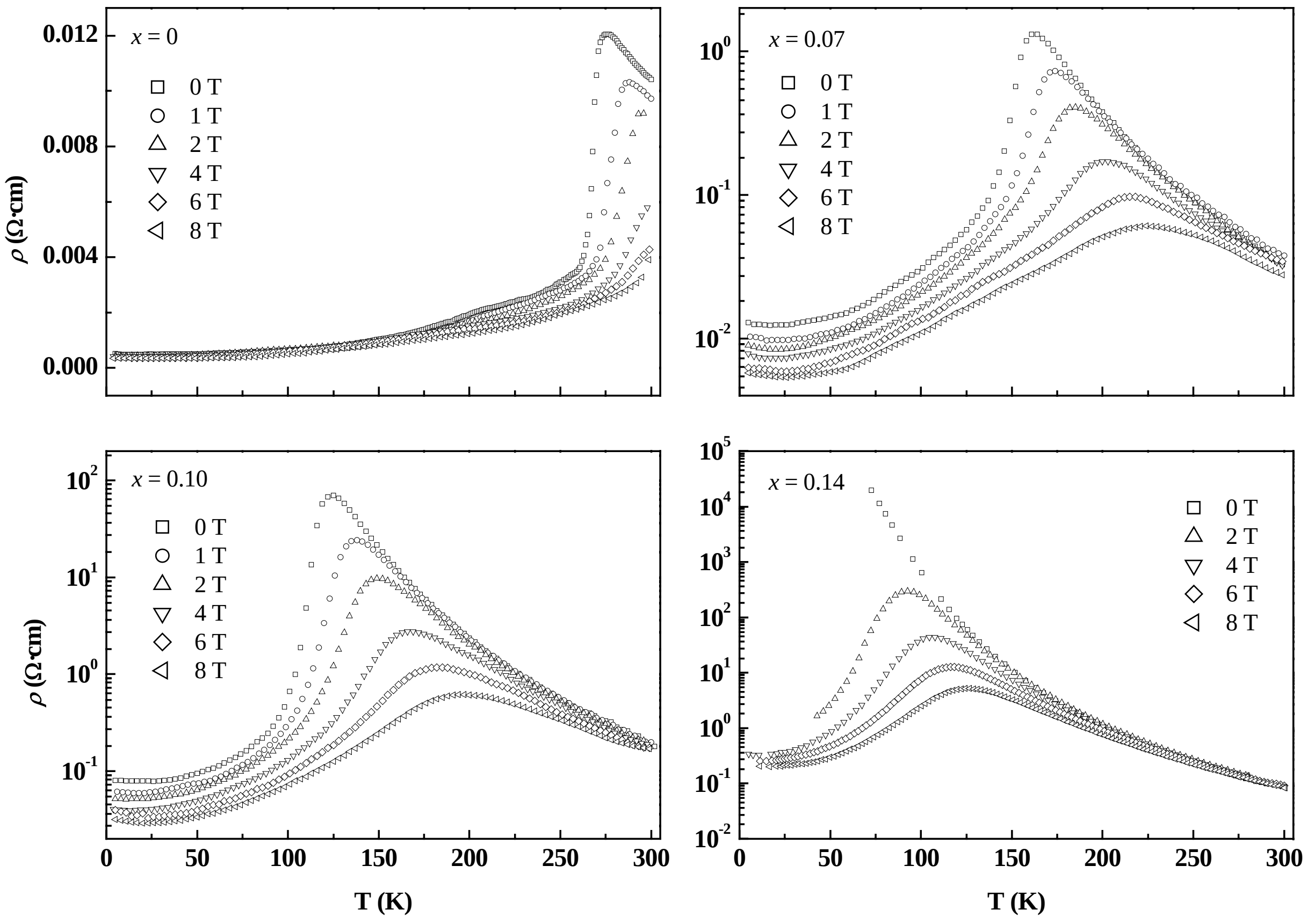

Fig. 2 by G. Li *et al.*

-

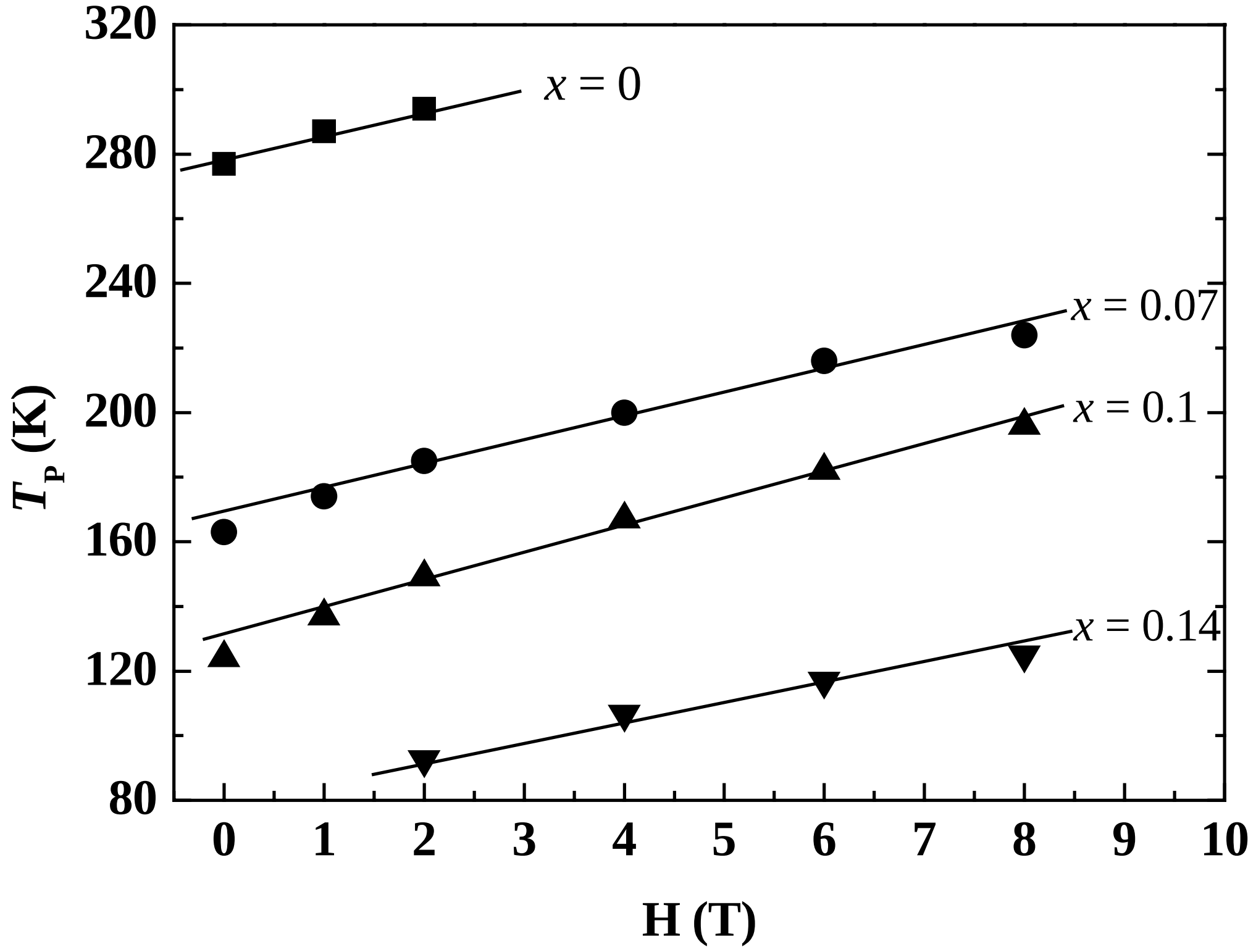

Fig. 3 by G. Li *et al.*

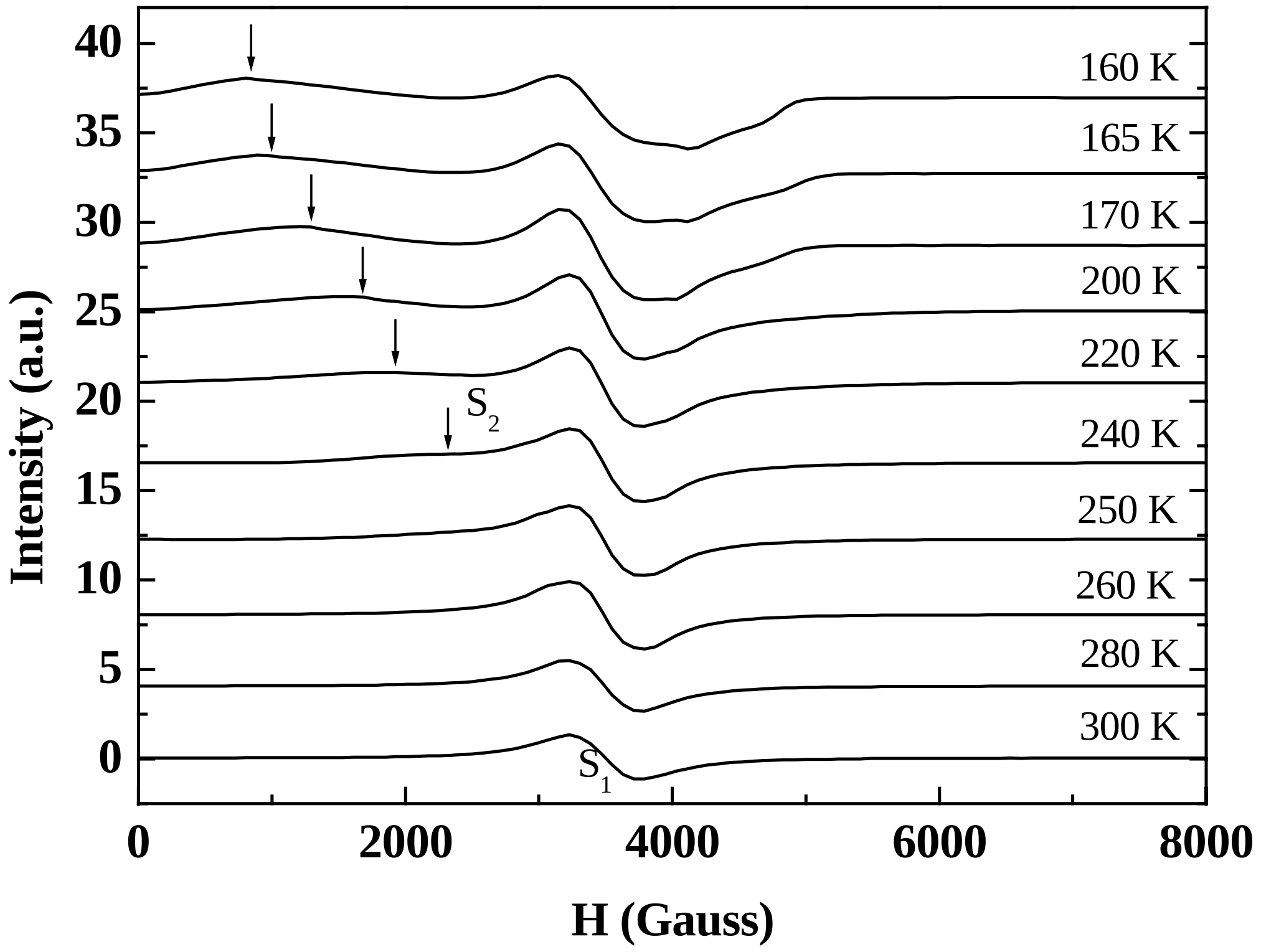

Fig. 4 by G. Li *et al.*

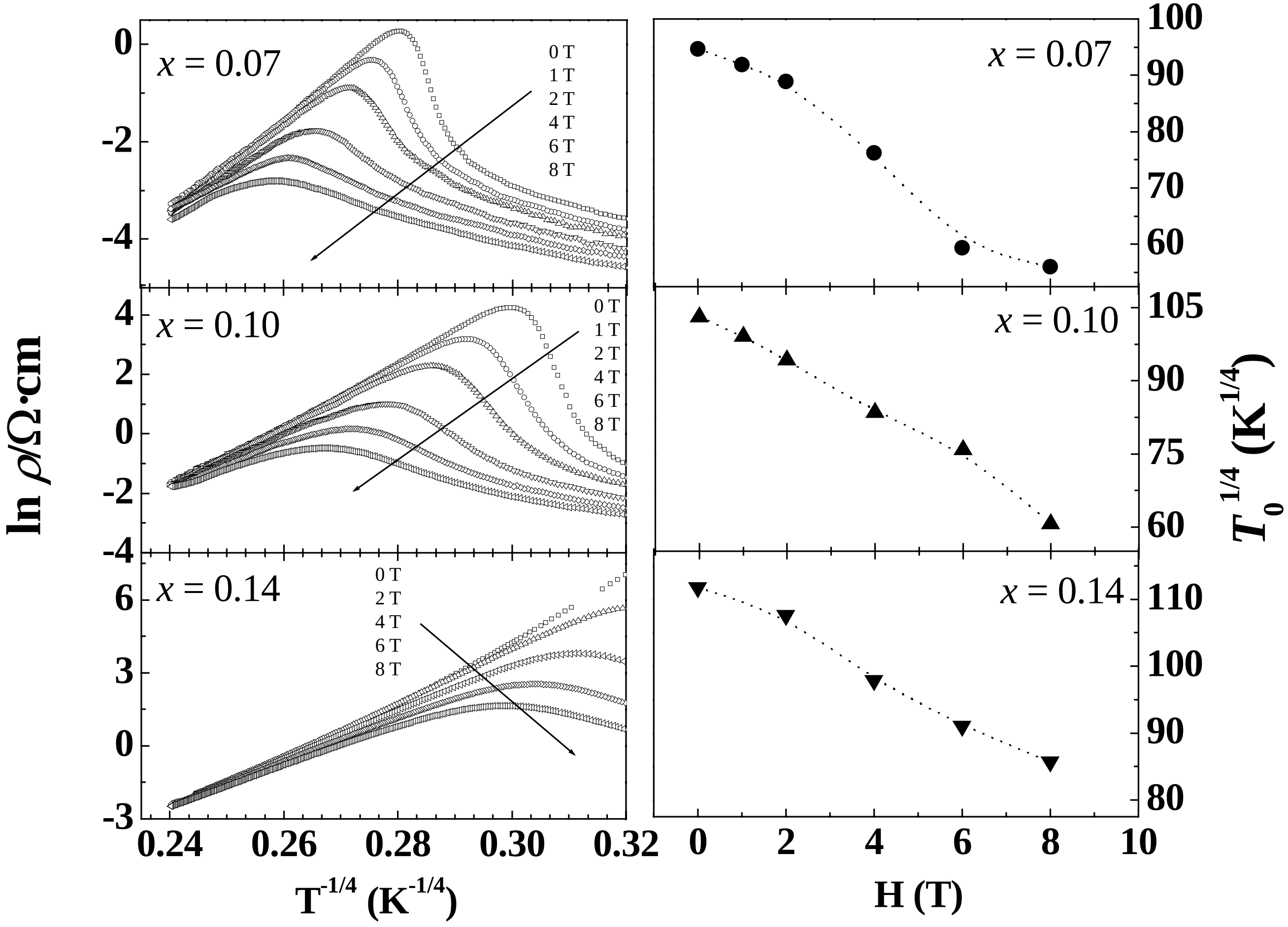

(a) (b)

Fig. 5 by G. Li *et al.*

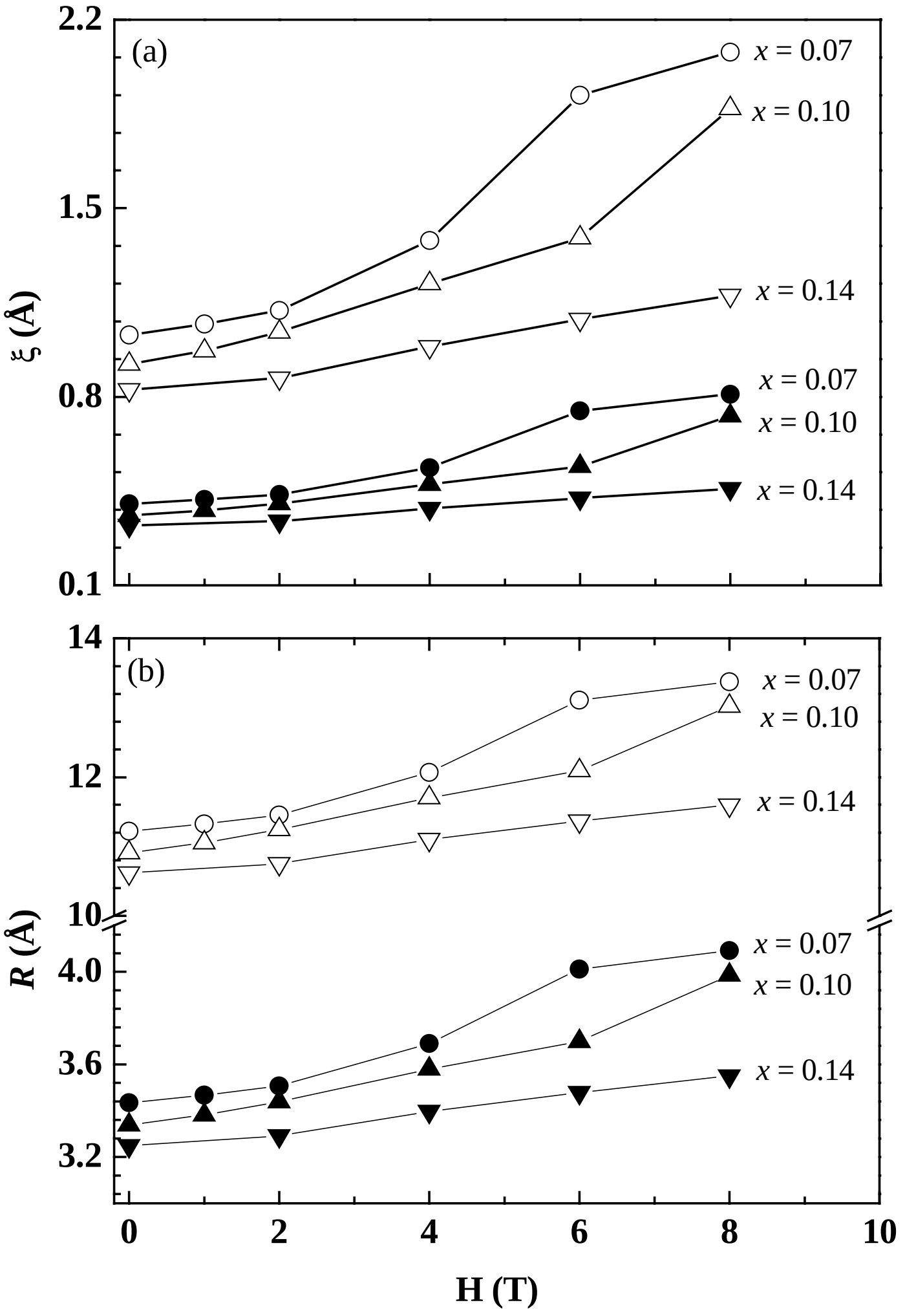

Fig. 6 by G. Li *et al.*